\documentclass[11pt, a4paper]{article}

\usepackage[english,francais]{babel}
\usepackage[OT1]{fontenc}

\usepackage{url}

\usepackage{hyperref}
\hypersetup{colorlinks=true}

\newcommand{\limitations}[1]{\newline {\it Limitations} --- #1}

\newcommand{\libtab}[7]
           {
             \begin{tabular}{|l|l|l|}
               \hline
               #1 & Distribution #2 -- #3 & Doc. updated in #4 \\
               \hline
               \multicolumn{3}{|l|}{\url{#5}}\\
               \hline
               \multicolumn{2}{|l|}{License: #6GPL compatible)} & Language: #7 \\
               \hline
             \end{tabular}
           }

\newcommand{\parag}[1]{\hspace*{\fill} \newline {\it #1 }}

\newcommand{\libdesc}[4]
{
  {\parag {Main features \newline} #3}\\
  {\parag {Portability: platforms and compilers supported \newline} #1}\\
  {\parag {Installation \newline} #2}\\
  {\parag {Limitations \newline} #4}
}

\title{Configuration File Parser Library}

\author{Claire Mouton - claire.mouton@inria.fr}

\begin{document}
\selectlanguage{english}

\maketitle
\tableofcontents

\newpage
\part*{Introduction}
The two basic methods to pass parameters to the main()-routine are: input
files and command line arguments. For small scale programs these input methods
allow to change parameters without having to recompile or having to create an
input file parser. Even for large programs input files and command line
arguments are a comfortable way to replace annoying graphical user
interfaces. When debugging complicated code, it is essential to be able to
isolate code fragments into a lonely main()-routine. In order to feed these
isolated code fragments with realistic data, sophisticated command line and
input ﬁle parsing becomes indispensable (excerpt from GetPot documentation).

\part{Requirements}

This document has been written to help in the choice of a configuration file
parser library to be included in Verdandi, a scientific library for data
assimilation. The main requirements are
\begin{enumerate}
\item Portability: Verdandi should compile on BSD systems, Linux, MacOS, Unix
  and Windows. Beyond the portability itself, this often ensures that most
  compilers will accept Verdandi. An obvious consequence is that all
  dependencies of Verdandi must be portable, especially the configuration file
  parser library.
\item The configuration file parser library should be written in C++, possibly
  in C, to be called from Verdandi written in C++.
\item License: any dependency must have a license compatible with Verdandi
  licenses (GPL and LGPL).
\item The library must provide the functionality of creating sections in the
  configuration file.
\end{enumerate}

\newpage
\part{Config}

\libtab {Config} {1} {May 2008} {May 2008}
{http://www.codeproject.com/KB/files/config-file-parser.aspx} {LGPL (} {C++}

\libdesc 
{Only standard C++ (code is cross platform) \\
  -- MSVS Express 2005 project files are included \\
  -- Manual build successfully tested with GNU g++/Linux}
{}
{-- Configuration files may be sub-structured arbitrarily deep \\
  -- Configuration files support the expansion of symbolic values from
  previously defined variables and environment variables}
{Sub groups less easy to create and read than with GetPot (requires an
  iterator on a STL map)}

\newpage
\part{GetPot}

\libtab {GetPot} {1.1.18} {Jul. 2008} {Mar. 2007}  {http://getpot.sourceforge.net/}
{LGPL (} {C++, Java, Python, Ruby}

\libdesc 
{Platform independent (istream problem with carriage return/newline on
  Microsoft Visual Studio should be solved)}
{-- Easy to install: contained in one single header file \\
  -- Requires STL library}
{ -- Parses comand line arguments, single or multiple configuration files \\
  -- Variables can be sorted into sections and subsections \\
  -- Many features, such as unrecognized object detection (but I found a
  bug...) and prefixes allowing to focus on variables in a given section \\
  -- Easy to use thanks to detailed, organized and commented example files
  (see: \url{http://getpot.sourceforge.net/example.html}) \\
  -- Provides a language allowing a variety of operations on variables,
  numbers and strings inside a configuration file (recursive replacements,
  conglomerate variable names, dictionaries; concatenation and replacement in
  strings; power expressions, comparisons and conditions in numeric
  expressions (see example file:
  \url{http://getpot.sourceforge.net/expand.html})) \\
  -- The user can define and name its own variables (the parser finds them) \\
  -- GetPot can be used to emulate trivial function calls (without syntax
  checking) \\
  -- Several styles of comment available for configuration files \\
  -- Emacs syntax highlighting for GetPot files \\
  -- Provides a Python port \\
  -- Non-military usage only, but civil applications are allowed even in a
  military context}
{-- No exceptions thrown (for example, no error message when a file is not
  found; when numeric expressions have a wrong type, refer to non-existent
  variables), a default value is returned if a key is not found in the input
  file \\
  -- Subsection definition is not intuitive and not flexible if './' syntax is used,
  better to use only full subsection names \\
  -- Header and source not fully separated}

\newpage
\part{libcfgparse}

\libtab {libcfgparse} {0.6} {Dec. 2005} {Dec. 2005}  {http://freshmeat.net/projects/libcfgparse/}
{GPL (} {C}

\libdesc 
{}
{Requires yacc and lex}
{-- C, C++ and Python API\\
  -- Configuration file syntax is similar to C \\
  -- Allows to keep data structured as lists and records; by this allows to
  group variables together to handle them conveniently \\
  -- Built-in variable types: string, boolean, integer and floating-point \\
  -- Allows pre-definitions of types and declaration of default values \\ 
  -- Allows for strict type-checking \\
  -- Allows to parse data directly from memory without a file}
{-- Preliminary documentation \\
  -- No maintenance, not in development \\
  -- Requires yacc and lex (so designed for Unix operating system) \\
  -- Written in C (but provides a C++ API)}

\newpage
\part{libconfig}
\libtab {libconfig} {1.3.2} {Feb. 2009} {Feb. 2009}  {http://www.hyperrealm.com/libconfig/}
{LGPL (LGPL and } {C++}

 \libdesc 
 {-- Linux, Solaris and Mac OS X \\
   -- Windows 2000/XP and later: gcc in the MinGW environment or with Visual
   Studio 2005}
  {}
  {-- A fully reentrant parser: independent configurations can be parsed in
    concurrent threads at the same time \\
   -- Both C and C++ bindings, as well as hooks to allow for the creation of
   wrappers in other languages \\
   -- A low-footprint implementation (just 38K for the C library and 66K for
   the C++ library) that is suitable for memory-constrained systems \\
   -- Proper documentation}
 {Configuration file syntax similar to C++ (quite heavy with \{\} and ; )}

\newpage
\part{libConfuse} 

\libtab {libConfuse} {2.6} {Dec. 2007} {2004}  {http://www.nongnu.org/confuse/}
{LGPL (} {C}

\libdesc 
{}
{}
{-- Supports sections and (lists of) values (strings, integers, floats,
  booleans or other sections), as well as some other features (such as
  single/double-quoted strings, environment variable expansion, functions and
  nested include statements)\\
  -- No myriads of features in a simple API making it easy to use and quick to
  integrate with one's code}
{Written in C}

\newpage
\part{Program\_options}

\libtab {Program\_options} {1.38.0} {?} {Nov. 2007}
{http://www.boost.org/doc/libs/1_38_0/doc/html/program_options.html} {Boost
  Software License (} {C++}

\libdesc 
{Any platform}
{Program\_options is one of the Boost C++ Libraries}
{-- Reads key/value pairs from command line, a configuration file and
  environment variables \\
  -- Good documentation and tutorial }
{}

\newpage
\part{RudeConfig™}
\libtab {RudeConfig™} {5.0.5} {released in ?} {?}  {http://rudeserver.com/config/}
{GPL (} {C++}

\libdesc 
{Borland, Visual C++ 6.0 and Linux}
{}
{-- Independent core library designed for CGI (Common Gateway Interface) \\
  -- Reads and writes configuration / .ini files \\
  -- Provides fully customizable delimiters and comment characters, ensuring
  compatibility with most existing configuration / .ini file formats \\
  -- Allows multiline values using backslash escapes \\
  -- Comments within the configuration file are fully preserved when the
  contents are re-saved  \\
  -- Deleted values can become commented out when the object is re-saved,
  preserving old data \\
  -- LGPL license can be provided by the author if needed}
{}

\newpage
\part{Talos}

\libtab {Talos} {1.0} {Apr. 2007} {Oct. 2004}  {http://vivienmallet.net/lib/talos/}
{GPL (} {C++}

\libdesc 
{Manages Windows end of line character}
{}
{Provides miscellaneous functions and objects, such as \\
  -- Functions to deal with strings and files that C++ is missing \\
  -- An extended ifstream class \\
  -- Two classes to read flexible configuration files: one for a single
  configuration file and another for multiple configuration files \\ \\  
  The parser functionalities are: \\
  -- Search for a word in the configuration file \\
  -- Extracts the value of a field \\
  -- Benefits from markup substitution \\
  -- Exceptions may be thrown providing a string explaining what happened \\
  -- Allows to apply a constraint on a value, or to force it to be part of a
  set of values \\
  -- Very clear and flexible configuration file syntax}
{No subsection}

% \libtab {Library name} {number} {...} {...}  {website}
% {GPL?, GPL compatible?} {language}
% \libdesc 
% {Linux}
% {Requires ...}
% {Users (examples)}
% {miscellaneous/license/language C++/portability/section}

\newpage
\part{Other Libraries}

%\paragraph{} : \url{}. \limitations {.}

\paragraph{A Configuration Package} :
\url{http://alumni.media.mit.edu/~rahimi/configuration/}. \limitations {Based
  on Lex and Yacc, no activity since 2001.}

\paragraph{ccl} 
(Customizable Configuration Library) allows the comment, key/value and string
literal delimiters to be programatically specified at runtime. Simple,
portable with a small interface consisting of five functions :
\url{http://sbooth.org/ccl/}. \limitations {Written in C.}

\paragraph{CFL}
(Configuration File Library). Portable, requires GDSL library:
\url{http://home.gna.org/cfl/}. \limitations {Written in C.}

\paragraph{ConfigFile} C++ portable code with templates:
\url{http://www-personal.umich.edu/~wagnerr/ConfigFile.html}. \limitations
{Impossible to define sections in configuration files.}

\paragraph{Configuration File Parser}
supports both shared as well as static binding of binaries. Provides API for
both manual and automatic processing of configuration file entries :
\url{http://sourceforge.net/projects/parser}. \limitations {Written in C.}

\paragraph{ConfigParser} 
reads and writes configuration files with a syntax similar to C/C++; after
being read in a configuration file, the configuration data is available for
access, modification and removal, or can also be written back to a file:
\url{http://www.codedread.com/code.php#Config}. \limitations {Not maintained,
  not portable (only Windows executable without source code).}

\paragraph{Confix} :
\url{http://www.flipcode.com/archives/Configuration_File_Parser.shtml}. \limitations
{IBM Public license not compatible with GPL license.}

\paragraph{dot.conf} : \url{http://www.azzit.de/dotconf/}. \limitations {Written in C, no
  activity since 2003.}

\paragraph{dotconfpp} : \url{http://ostatic.com/dotconfpp/}. \limitations {Not portable (distribution FreeBSD for i386).}

\paragraph{GLib} 
is a general-purpose utility library, which provides many useful data types,
macros, type conversions, string utilities, file utilities, a main loop
abstraction, and so on. It works on many UNIX-like platforms, Windows, OS/2
and BeOS. LGPL license. Provides a key-value file parser:
\url{http://library.gnome.org/devel/glib/unstable/glib-Key-value-file-parser.html}. 
\limitations {Written in C.}

\paragraph{iniParser}
Simple, small, portable and robust ini file parsing library.:
\url{http://ndevilla.free.fr/iniparser/}. \limitations {Written in C.}

\paragraph{Libconfig} supports callback functions, automatic variable
assignment : \url{http://www.rkeene.org/oss/libconfig}. \limitations {Written
  in C, not tested under Windows.}

\paragraph{libinifile} :
\url{http://it.bmc.uu.se/andlov/proj/libinifile/}. \limitations {Not portable
  (uses make to build the library under Unix).}

\paragraph{liblcfg}
is a lightweight configuration file library. The file format supports
arbitrarily nested simple assignments, lists and maps (aka dictionaries):
\url{http://liblcfg.carnivore.it/}. \limitations {Written in C99.}

\paragraph{ParseCfg} 
: \url{http://www.mentaljynx.com/CCpp/libs/parsecfg/}. \limitations {Written
  in C, no good documentation, no activity since 2001.}

\paragraph{spConfig} 
uses configuration files with an XML-like syntax with some additional
preprocessor-style commands:
\url{http://prj.softpixel.com/spconfig/}. \limitations {Written in C, no
  activity since 2003.}

\paragraph{Templatized Configuration File Parser (TConf)} 
is inspired by the TCLAP project (Templatized C++ Command Line Parser Library:
\url{http://tclap.sourceforge.net/}). TConf is a smaller configuration parser
library and can be used in combination with TCLAP. Released in Aug. 2007 under
GPL license: \url{http://tconf.sourceforge.net/}. \limitations {No sections in
  configuration files.}

\paragraph{TCFP}
(Tiny Config File Parser library) is small, fast and simple:
\url{http://sourceforge.net/projects/tcfp/}. \limitations {Pre-alpha version,
  written in C99.}

%\newpage
%\section*{Acknowledgement}
%This document benefits from discussions with Vivien Mallet.

\end{document}